\documentclass{article}
\usepackage{epsfig}
\def \pt   {p_{\rm\scriptscriptstyle T}}

\newcommand\ep{\epsilon}
\newlength{\capwidth}
\begin{document}
\title{
\begin{flushright}
 \vspace*{-1.5in}
\begin{minipage}{3cm}
 {\normalsize Bicocca-FT-03-1\\[-2ex]
 January 2003 \\[-2ex]
hep-ph/0301003}
\end{minipage}
\end{flushright}
\vspace*{0.5in}
HEAVY FLAVOUR PRODUCTION
\footnote{Invited review talk at the
First International Workshop on Frontier Science, ``Charm, Beauty and CP'',
October 6-11, 2002, INFN Laboratory, Frascati, Italy.}
}
\author{
P. Nason \\
{\em INFN, Sez. di Milano, Milan, Italy}
}
\maketitle
\begin{abstract}
I discuss few recent developments in Heavy Flavour Production phenomenology.
\end{abstract}
\section{Introduction}
The phenomenology of Heavy flavour production attracts considerable
theoretical and experimental interest.

The theoretical framework for the description of heavy flavour production is
the QCD improved parton model. Besides the well-established NLO corrections to
the inclusive production of heavy quark in hadron-hadron\cite{Nason:1988xz,
Nason:1989zy,Beenakker:1991ma,Mangano:1992jk},
hadron-photon\cite{Ellis:1989sb,Smith:1992pw,Frixione:1994dg}, and
photon-photon collisions\cite{Drees:1993eh}, much theoretical work has been
done in the resummation of contributions enhanced in certain regions of phase
space: the Sudakov region, the large transverse momentum region and the
small-$x$ region.

The theoretical effort involved is justified by the large variety of
applications that heavy quark production physics has, in top, bottom and charm
production.  Besides the need of modeling these processes, heavy quark
production is an important benchmark for testing QCD and parton model ideas,
due to the relative complexity of the production process, the large range of
masses available, and the existence of different production environments, like
$e^+e^-$ annihilation, hadron-hadron, photon-hadron and photon-photon
collisions.  Although the order of magnitude of the total cross sections, and
the shape of differential distribution is reasonably predicted, in some areas
large discrepancy are present, especially for $b$ production.

\section{Total cross section for top and bottom}
\begin{figure}[htb]
\begin{center}
\epsfig{file=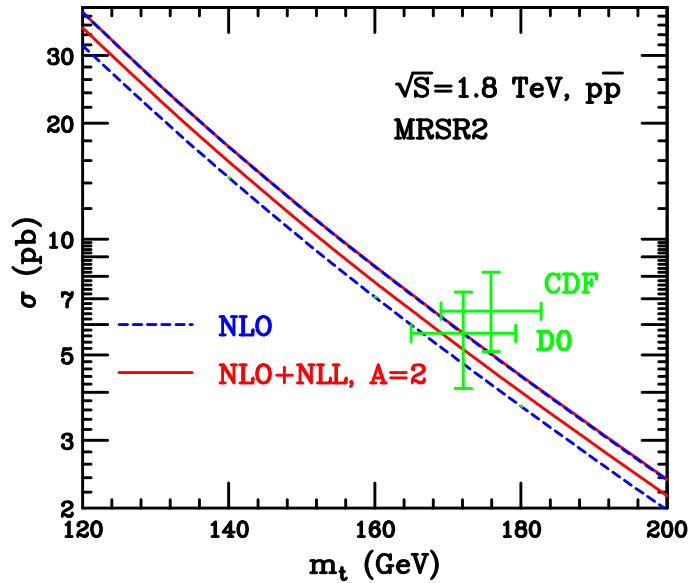,width=0.6\textwidth}
\end{center}
 \caption{\it
    Results on top cross sections at the Tevatron.
    \label{topxsec} }
\end{figure}

Top production\cite{Affolder:2001wd,Abazov:2002gy}
has been a most remarkable success of the theoretical
model. The measured cross section has been found in good agreement
with theoretical calculations, as shown in fig.~\ref{topxsec}.
Resummation of soft gluon effects\cite{Bonciani:1998vc} reduces the
theoretical uncertainty in the cross section, pushing it toward the
high side of the theoretical band. It remains, however, inside
the theoretical band of the fixed order calculation, thus showing
consistency with the estimated error.

Recently, the HERA-B experiment has provided a new
measurement of the $b\bar{b}$ total
cross section\cite{Abt:2002rd}. Their result is in good agreement
with previous findings\cite{Jansen:1995bz,Alexopoulos:1997zx}.
More details on this measurement have been given in Saxon's talk\cite{saxon}.
Here I will only make some remarks on the theoretical aspects.
This experiment is sensitive to the
moderate transverse momentum region, where the bulk of the total cross
section is concentrated. Since the production
is (in a certain sense) close to threshold, resummation of Sudakov effects
is important also in this case, and brings about a considerable
reduction of the theoretical uncertainty\cite{Bonciani:1998vc}.
The HERA-b result is compatible with the central value prediction,
with the $b$ pole mass around
$4.75$ GeV. Higher precision may constrain further the $b$ quark mass.
\section{Differential distributions}
The Tevatron has had a longstanding disagreement with QCD in
the $b$ transverse momentum spectrum. 
A recent publication of the $B^+$ differential cross section
by CDF\cite{Acosta:2001rz} has quantified
the disagreement as a factor of $2.9 \pm 0.2 \pm 0.4$ in the ratio
of the measured cross section over the theoretical prediction.
This discrepancy has been present since a long time,
and it has been observed both in CDF and D0.
Some authors\cite{Berger:2000mp} have argued that the discrepancy could be
interpreted as a signal for Supersimmetry.

Because of the large theoretical uncertainties, this discrepancy has been often
downplayed. In fact, several effects may conspire to raise the $b$ cross
section to an appropriate value: small-$x$
effects\cite{Nason:1988xz,Catani:1991eg,Collins:1991ty},
threshold effects\cite{Bonciani:1998vc} and
resummation of large $\log \pt$. It is not however clear whether these effect
can be added up without overcounting. Furthermore, they are all higher
order effects, and thus should not push the cross section too far out
of the theoretical band, which includes estimates of unknown higher
order effects.

Recently, a small-$x$
formalism\cite{Ciafaloni:1988ur,Catani:1990yc,Catani:1990sg,Marchesini:1995wr}
has been
implemented in a Monte Carlo program\cite{Jung:2001hx}
(the CASCADE Monte Carlo),
and it has been claimed that this programs correctly predicts the $b$ spectrum
at the Tevatron\cite{Jung:2001rp}.
Although encouraging, this result should be regarded with a
word of caution. The formalism involved only accounts for leading small $x$
effects, and does not correct for the lack of leading terms, that are known to
be important for heavy flavour production at the Tevatron regime.
Experience with other contexts where resummation techniques have been
applied has taught us that it is not difficult to overestimate the
importance of resummation effects, and much study is needed to
reliably assess their importance\cite{Catani:1996yz}.

It has been observed since some time that an improper understanding of
fragmentation effects may be one of the causes of the Tevatron
discrepancy. This possibility stems from the fact that $b$ quark cross
sections are in reasonable agreement with the Tevatron measurements of
the $B$ meson spectrum, while the cross section obtained by applying a
standard fragmentation function of the Peterson
form\cite{Peterson:1983ak} with $\ep=0.006$ to the quark cross section
yields too soft a spectrum.  It was suggested\cite{Frixione:1997nh} to
study $b$ quark jets rather than $B$ meson's distributions.  In fact,
the jet momentum should be less sensitive to fragmentation effects
than the hadron momentum. A D0 study \cite{Abbott:2000iv} has
demonstrated that by considering $b$ jets instead of $B$ hadrons the
agreement between theory and data improves considerably.

It was recently shown\cite{Cacciari:2002pa} that an accurate assessment
of fragmentation effects brings about a reduction of the discrepancy
from the factor of 2.9 quoted by CDF to a factor of 1.7.
This remarkable reduction is essentially due to recent progress
in fragmentation function measurements by LEP experiments and the
SLD\cite{Heister:2001jg,Abe:2002iq,Abbiendi:2002vt,delphi}
(the current status of fragmentation measurements
has been reviewed in Boccali's talk\cite{boccali}),
and by a particular method for extracting
the relevant information about non-perturbative fragmentation
effects from the $e^+e^-$ data. Here I will not enter in the
details of the method, that have been described in\cite{Cacciari:2002pa}.
I will instead try to give an overall illustration of the main
features of the method.
First of all, strictly speaking, the description of fragmentation
effects in terms of a fragmentation function (i.e., a probability
distribution for an initial quark to hadronize into a hadron
with a given fraction of its momentum) is only valid at very large
transverse momenta. One could then extract the fragmentation function
from LEP data, and use it in high $\pt$ $B$ production at the Tevatron.
Unfortunately, the regime of large transverse momenta is not quite
reached at the Tevatron.
For example, the differential cross section $d\sigma/d\pt^2dy$
at the Tevatron, for $\pt=10,y=0$, computed at the NLO level
including mass effects is $12.1\,{\rm nb/GeV^2}$, while in the massless
approximation (i.e., neglecting terms suppressed as powers of $m/\pt$)
is $1.78\,{\rm nb/GeV^2}$. At $\pt=20$ we have $0.372\,{\rm nb/GeV^2}$
for the full massive, and $0.220\,{\rm nb/GeV^2}$ for the massless
limit, which starts to approach the asymptotic value. As a rule of thumb,
the massless approximation starts to approach the massive one
around $\pt \approx 5 m$\footnote{This elementary fact has been known
for more than five years, although, surprisingly enough,
some authors prefer to ignore it even now\cite{Kniehl:2002xn}.}.
It does therefore make no sense to use any massless approach
for $\pt < 5 m$.
In earlier work on heavy quark production at large transverse
momentum\cite{Cacciari:1994mq}, this fact went unnoticed,
since large higher order terms (i.e. beyond the NLO level)
in the fragmentation function approach accidentally
compensated for the lack of mass terms,
thus giving the impression that the massless approach is good
down to $\pt\approx 2m$.

\begin{figure}[htb]
\begin{center}
\epsfig{file=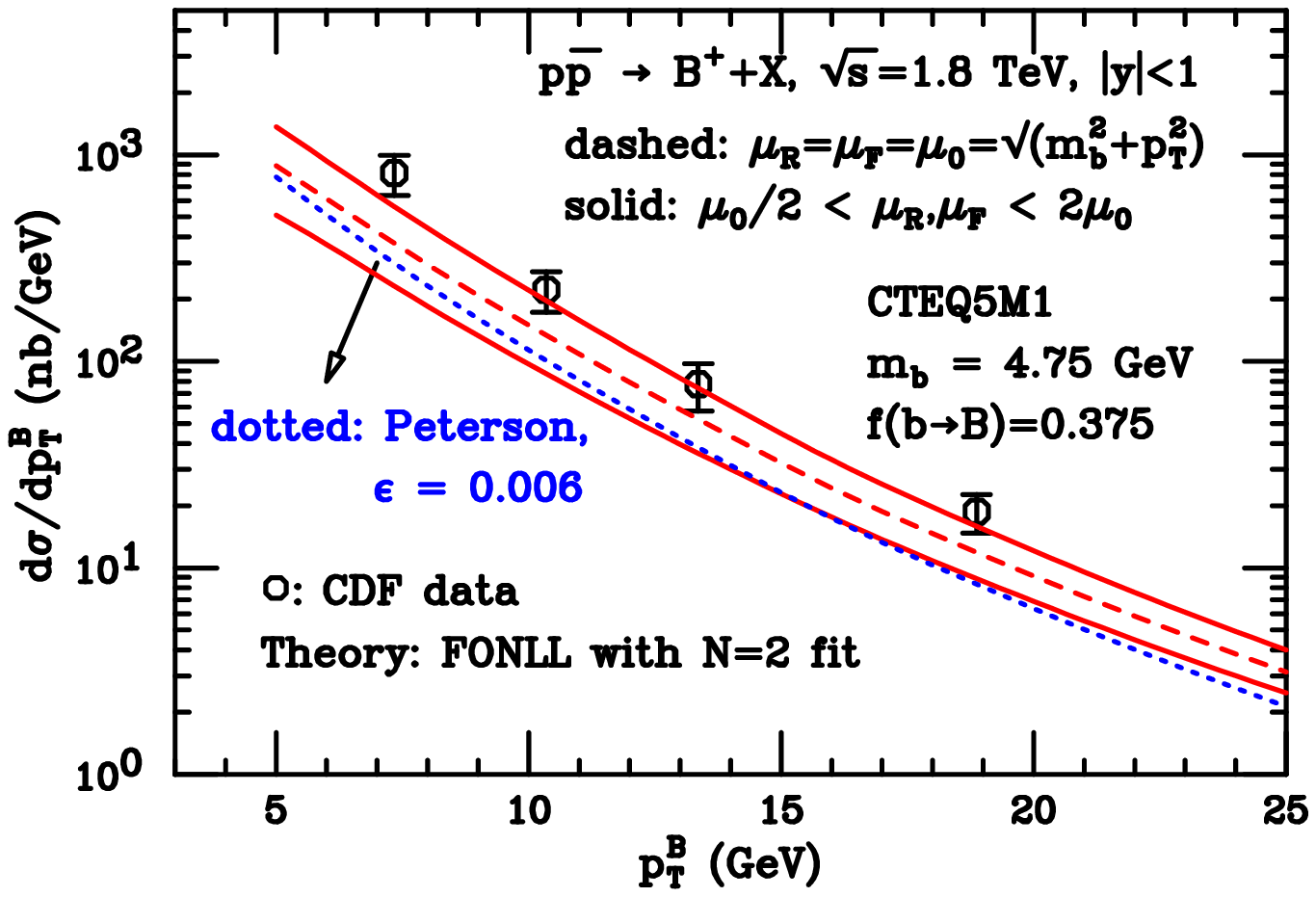,width=0.55\textwidth}
\epsfig{file=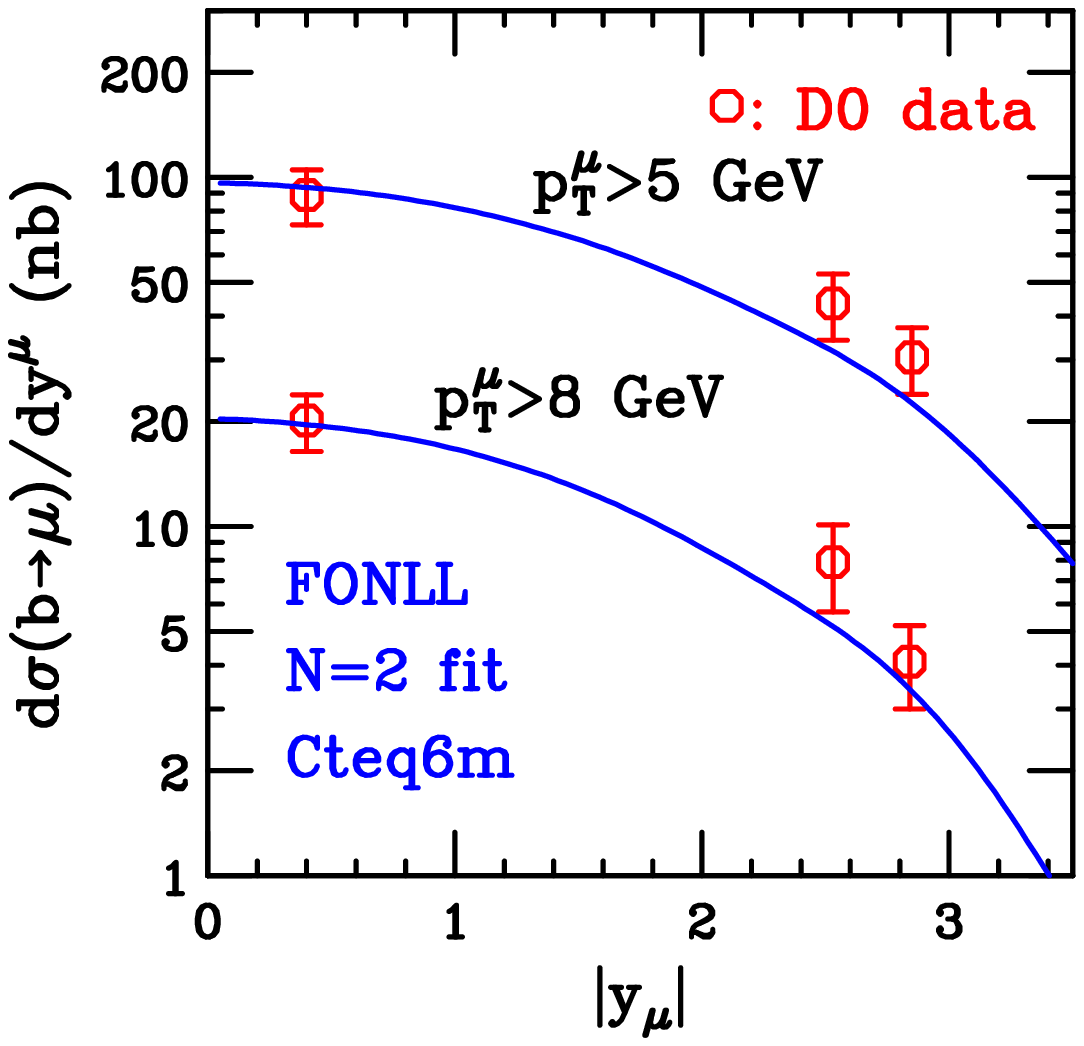,width=0.40\textwidth}
\end{center}
 \caption{\it
Comparison of CDF\cite{Acosta:2001rz} and D0\cite{Abbott:1999wu} data with
the FONLL calculation\cite{Cacciari:2002pa}.
 }
\label{Bhadcn}
\end{figure}

In order to perform a calculation
that is reliable in both the low and the high transverse momentum
regime, we have thus to merge the massive NLO calculation
with the fragmentation function approach. The merging point
must therefore be around $pt\approx 5m$.
A summary of the theoretical tools that have lead to the matched
(so called FONLL) approach are summarised as follows:
\begin{itemize}
\item[1] Single inclusive particle production in hadronic
collisions\cite{Aversa:1990uv}. Single hadron
production are described in term of NLO single parton cross section
convoluted with a NLL fragmentation function;
\item[2] Heavy quark Fragmentation Function\cite{Mele:1991cw};
a method for the computation of the heavy quark fragmentation function at
all orders in perturbation theory is developed, and applied at NLL.
Several applications in $e^+e^-$ physics have appeared
\cite{Colangelo:1992kh,Cacciari:1997wr,Cacciari:1997du,Nason:1999zj}.
\item[3] Single inclusive heavy quark production at large
$p_T$\cite{Cacciari:1994mq}; item 1 and 2 are combined to give
a NLL resummation of transverse momentum logarithms in heavy quark production;
\item[4] FONLL calculation of single inclusive heavy quark production;
item 3 is merged without overcounting with standard NLO
calculations. This procedure has been implemented both in
hadroproduction\cite{Cacciari:1998it} and
in photoproduction\cite{Cacciari:2001td,Frixione:2002zv}.
\end{itemize}
A summary of comparison of the FONLL calculation with data is given in
fig.~\ref{Bhadcn}. More details on the CDF measurement have been
given in Saxon's talk. The comparison with D0 data (still preliminary)
shows very good agreement, compared to the discrepancy of a factor
of order 3 found in the D0 publication.
\section{Conclusions}
The theory of heavy flavour production seems to give a good
qualitative description of the available data.
In the case of top production, the comparison between theory and experiment
is satisfactory also at a quantitative level.

Recent progress has taken place in the field of $b$ hadroproduction.
The HERA-b experiment has provided a cross section for $b$ production
at relatively low CM energy. Some progress in understanding the role
of fragmentation has considerably reduced the longstanding problem
of the $b$ momentum spectrum at the Tevatron.

Major problems do remain in the
(perhaps less developed) areas of bottom production in $\gamma\gamma$
and $\gamma p$ collisions. Discussion of these problems were reported
in Saxon's and Achard's talk in this conference\cite{saxon,achard}.
In both contexts, while charm production seems to be reasonably well
described by QCD, there is an excess of bottom production, which
seems to be far out of the theoretical uncertainty band.
These problems have been around for sometimes now. On the
positive side, the recent Zeus measurements presented in\cite{saxon}
seem to show a smaller discrepancy than previously found.

\providecommand{\href}[2]{#2}\begingroup\raggedright\endgroup
\end{document}